\documentclass[twocolumn,showpacs,superscriptaddress,
               preprintnumbers,amsmath,amssymb,floatfix,prl]{revtex4}

\usepackage{graphicx}
\usepackage{dcolumn}
\usepackage{bm}

\newcommand{\prot}{\rm p}
\newcommand{\elec}{\rm e}
\newcommand{\mup}{\ensuremath{\mu \prot}}

\newcommand{\pip}{\ensuremath{\pi \prot}}
\newcommand{\xray}{X-ray}
\newcommand{\xrays}{X-rays}

\newcommand{\Htwo}{\ensuremath{\mathrm{H_{2}}}}
\newcommand{\mus}{\ensuremath{\mu{\rm s}}}
\newcommand{\OneS}{\ensuremath{1S}}
\newcommand{\TwoS}{\ensuremath{2S}}
\newcommand{\TwoP}{\ensuremath{2P}}

\newcommand{\muAu}{\ensuremath{\mu \mathrm{Au}}}
\newcommand{\muXe}{\ensuremath{\mu \mathrm{Xe}}}
\newcommand{\ppmu}{\ensuremath{(\prot \prot \mu)^+}}
\newcommand{\ppmustar}{\ensuremath{ [(\prot \prot \mu)^+]^*}}
\newcommand{\ppmupee}{\ensuremath
                      {\{ [(\prot \prot \mu)^+]^*\,\prot \elec \elec \}^*}}
\newcommand{\EpsTS}{\ensuremath{\varepsilon_{\TwoS}}}

\newcommand{\EpsTSlong}{\ensuremath{\varepsilon_{\TwoS}^{\rm long}}}

\newcommand{\TauTSshort}{\ensuremath{\tau_{\TwoS}^{\rm short}}}
\newcommand{\TauTSlong}{\ensuremath{\tau_{\TwoS}^{\rm long}}}
\newcommand{\AmpTS}{\ensuremath{A_{\TwoS}}}
\newcommand{\OffTS}{\ensuremath{\Delta t_{\TwoS{}}}}
\newcommand{\LamTSquench}{\ensuremath{\lambda_{\TwoS{}}^{\rm quench}}}
\newcommand{\TauTSquench}{\ensuremath{\tau_{\TwoS}^{\rm quench}}}

\newcommand{\press}[1]{\ensuremath{p_{\rm #1}}}
\newcommand{\pHtwo}{\press{\Htwo}}
\newcommand{\ChiSq}{\ensuremath{\chi^2}}

\begin{document}


\title{Observation of Long-Lived Muonic Hydrogen in the \boldmath{2S}
  State}

\author{Randolf~Pohl}
\altaffiliation[Now at ]{MPI f\"ur Quantenoptik, Garching, Germany}
\email{Randolf.Pohl@mpq.mpg.de}
\affiliation{Institut f\"ur Teilchenphysik, ETH Z\"urich, 8093
  Z\"urich, Switzerland}
\affiliation{Paul Scherrer Institut, 5232 Villigen-PSI, Switzerland}
\author{Herbert~Daniel}
\affiliation{Physik Department, Technische Universit\"at M\"unchen,
  85748 Garching, Germany}
\author{F.~Joachim~Hartmann}
\affiliation{Physik Department, Technische Universit\"at M\"unchen,
  85748 Garching, Germany}
\author{Peter~Hauser}
\affiliation{Paul Scherrer Institut, 5232 Villigen-PSI, Switzerland}
\author{Franz~Kottmann}
\affiliation{Institut f\"ur Teilchenphysik, ETH Z\"urich, 8093
  Z\"urich, Switzerland}
\author{Valery~E.~Markushin}
\affiliation{Paul Scherrer Institut, 5232 Villigen-PSI, Switzerland}
\author{Markus~M\"uhlbauer}
\affiliation{Physik Department, Technische Universit\"at M\"unchen,
  85748 Garching, Germany}
\author{Claude~Petitjean}
\affiliation{Paul Scherrer Institut, 5232 Villigen-PSI, Switzerland}
\author{Wolfgang~Schott}
\affiliation{Physik Department, Technische Universit\"at M\"unchen,
  85748 Garching, Germany}
\author{David~Taqqu}
\affiliation{Paul Scherrer Institut, 5232 Villigen-PSI, Switzerland}
\author{Peter~Wojciechowski-Grosshauser}
\affiliation{Physik Department, Technische Universit\"at M\"unchen,
  85748 Garching, Germany}
%
%
\date{July 14, 2006}  

\begin{abstract}
The kinetic energy distribution of ground state muonic hydrogen
atoms \mup(\OneS) is determined from time-of-flight 
spectra measured at 4, 16, and 64\;hPa \Htwo{} room-temperature gas.
A 0.9\,keV-component is discovered and attributed to radiationless
deexcitation of long-lived \mup(\TwoS) atoms in collisions with
\Htwo{} molecules.
The analysis reveals a relative population of about 1\,\%, and a
pressure-dependent lifetime (e.g.\ $30.4\,^{+21.4}_{-9.7}$\;ns at
64\;hPa) of the long-lived \mup(\TwoS) population, 
equivalent to a \TwoS-quench rate in \mup(\TwoS)+\Htwo{} collisions of
$ 4.4\,^{+2.1}_{-1.8} \times 10^{11}\;{\rm s}^{-1}$ at liquid hydrogen
density.
\end{abstract}


\maketitle
%
%
A measurement of the Lamb shift in muonic hydrogen, i.e.\ the energy
difference of 0.2\;eV between the \TwoP- and \TwoS-states of $\mu^-$p
atoms~\cite{PACHU96,EIDES01,BORIE05}, is in progress at the
Paul Scherrer Institute (PSI), Switzerland~\cite{POHL05}. 
Vacuum polarization shifts the \TwoS{}-levels by 0.2\;eV {\it below} 
the \TwoP{}-levels; fine and hyperfine splittings of the $n=2$ levels 
are much smaller. The finite size effect is 2\,\% of the Lamb shift.

The \mup{} Lamb shift experiment
is expected to give a precise value for the
root-mean-square charge radius of the proton~\cite{PACHU99}. Together
with recent advances in H-atom spectroscopy~\cite{BEAU00, NIER00}, this
leads to a better determination of the Rydberg constant,
and to a test of bound-state quantum electrodynamics on a new level of
precision~\cite{MOHR05}.

The most important prerequisite for such an experiment, the
availability of sufficiently long-lived \mup(\TwoS) atoms,
has so far not been experimentally established.
When muons are stopped in \Htwo{} gas, \mup{} atoms are formed at high
$n$-levels and then deexcite predominantly to the ground state
(``muonic cascade'').
A fraction \EpsTS{} of a few percent reaches the metastable \TwoS{} state
whose lifetime is, in absence of collisions, essentially given
by the muon lifetime of 2.2\;\mus. 

In a gas, there is
collisional \TwoS-quenching, with very different rates depending on
the \mup{} kinetic energy being above or below the
\TwoS-\TwoP{} threshold ($\approx\!0.3$\;eV in the lab
frame)~\cite{JENSEN02}.
Most \mup(\TwoS) atoms are formed at energies above this
threshold~\cite{POHL00}, where collisional
\TwoS\,$\rightarrow$\,\TwoP{} Stark transitions (followed by
\TwoP\,$\rightarrow$\,\OneS{} radiative deexcitation) lead to rather
fast \TwoS-depletion. This is the ``short-lived''
  \TwoS-component with a predicted lifetime~\cite{CARFIO77}
$\TauTSshort \sim 100$\,ns/\pHtwo[hPa].
There is, however, a competition between such Stark transitions and 
deceleration~\cite{JENSEN02}.
A fraction of the \mup(\TwoS) atoms should therefore
survive the process of slowing down below 0.3\;eV, where 
transitions to the \TwoP{} state are energetically forbidden.
These \mup(\TwoS) atoms form the ``long-lived'' \TwoS-component, with
a lifetime \TauTSlong and a  
population \EpsTSlong{} (per \mup{} atom).

The \TwoS-population \EpsTS{} is well determined from the measured
\mup{} \xray{} yields~\cite{EGAN81,ANDE84,BRE98}.
\EpsTSlong can be derived from the
measured \OneS{} kinetic energy distributions~\cite{POHL01}
by using calculated elastic and inelastic \mup(\TwoS) cross
sections~\cite{JENSEN02}. At 16\;hPa, for example,
\EpsTS{}\,=\,$(4.40\pm 0.17)$\,\% and \EpsTSlong{}\,=\,$(1.16\pm 0.12)$\,\%.
Calculations of the radiative quenching process {\it during}
collisions~\cite{MUELLER75,COHEN81,MEN'SHIKOV86} predicted values of
$\TauTSlong{} \sim 600$\,ns or larger for an \Htwo{} pressure of
64\,hPa, but this could not be confirmed experimentally in searches
for delayed $n$\,=\,2$\rightarrow$1
\xrays{}~\cite{ANDE77,EGAN81,BOEC82}.

We report here on the first measurement of a non-radiative
deexcitation mechanism for long-lived \mup(\TwoS) atoms at 
gas pressures \pHtwo=\,4, 16, and 64\;hPa at room
temperature~\cite{POHL01}.
It was found in the analysis of time-of-flight (TOF) spectra
which were measured to determine the
\mup(\OneS) kinetic energy distributions at low densities.
A pronounced fast component showed up, corresponding to a 
surprisingly high \mup(\OneS) kinetic energy of 0.9\;keV.
%

%
%

The basic idea of the experiment is to form \mup(\OneS) atoms near the
axis of a cylindrical gas target and to deduce their kinetic energy
from the measured TOF they need to reach the gold-coated cylinder
surface.
The muon transfer reaction \mup\,+Au\,$\rightarrow$\,$\muAu^*$+p to
highly excited $\muAu^*$ takes place, and several \muAu{}
\xrays{} with energies up to 9\;MeV are emitted and detected by a
CsI(Tl) scintillator.
The cylindrical geometry gives a better kinetic energy resolution than
earlier planar geometries~\cite{ABBO97}.

\begin{figure}[t]
\includegraphics[width=0.9\columnwidth]{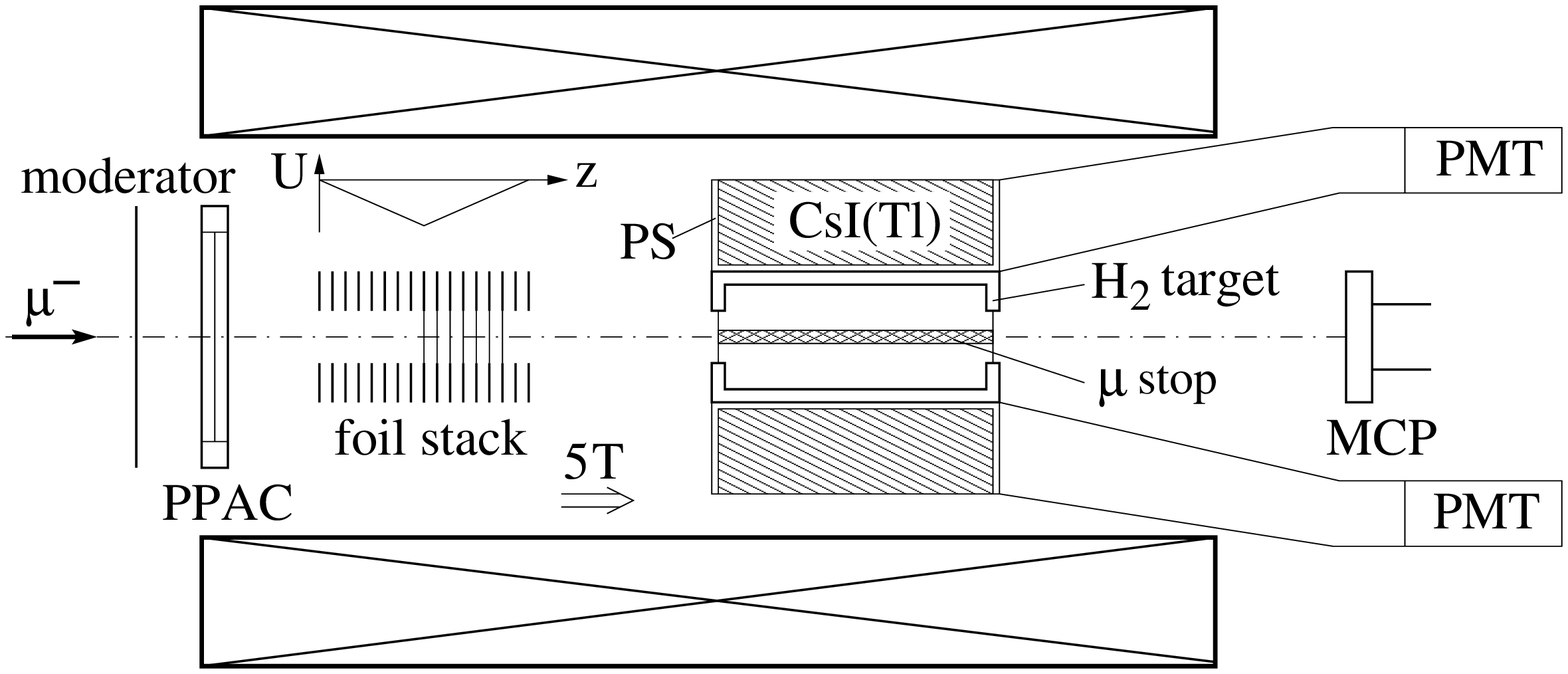}
  \caption{\label{fig:setup} Experimental setup inside the 5\;T solenoid
    (not to scale, dimensions given in text).
    PPAC: parallel plate avalanche counter; PS: plastic scintillator;
    MCP: microchannel plate; PMT: photomultiplier tube.  
    $U(z)$ is the electric potential of the 23 stack rings
    (7 with C-foils).
    }
\end{figure}

Figure~\ref{fig:setup} shows the experimental setup.  Negative muons
with a momentum of \,$p$\,=\,10\;MeV/c 
from the high-intensity $\pi$E5 channel at PSI
entered a 1\;m long, 20\;cm inner diameter solenoid operated at 5\;T
and evacuated to $10^{-5}$~hPa.
Its fringe field reduced the radial beam size to
a few mm in the region of full field.
About $5000\;\mu^-$\!/s reached a 16\;$\mu$m thick Mylar
entrance moderator.
The muons were detected by a parallel-plate avalanche counter (PPAC)
with a total areal density of 300\;$\mu$g/cm$^2$.
The PPAC is a double gas chamber filled with 16\;hPa isobutane, with a
detection efficiency up to 80\%, depending on the muon energy.
At the exit of the PPAC, the muons had energies between 0 and 200\;keV.
They then crossed a stack of seven thin carbon foils 
(each with an areal density of 5\;$\mu$g/cm$^2$ and 2\;cm diameter)
where a few electrons were released.  A high voltage applied to the
stack ($\Delta U$\,=\,1800\;V between successive foils) accelerated
these electrons, and frictional cooling~\cite{MUL99} enhanced the muon 
stop rate.
After traversing the gas target, the electrons were
detected in a microchannel plate (MCP).  
An incoming muon was identified by a delayed coincidence between the
signals from PPAC and MCP. 
The muon detection efficiency of the stack-MCP assembly was
$\sim\!40$\,\% for a target gas pressure of 4\;hPa and decreased to
$\sim\!15$\,\% at 64\;hPa.
The radial beam distribution was measured to be approximately
Gaussian-shaped with spreads
$\sigma_x \approx \sigma_y \approx 1.8$\;mm in the 5\;T field.
The cyclotron radii of the muons were $\sim$\,1\;mm at 5\;T.
%

%
%
%

The gas target vessel was a 26\;cm long Al cylinder.
Its inner cylindrical surface was coated with a 500\;nm thick 
evaporated gold layer.
The side walls were covered with polypropylene.  Measurements were
performed with two target inner radii, $R$\,=\,29\;mm and 10\;mm.
A target with polypropylene instead of gold was used to
determine the background.
Muon transfer to carbon does not produce \xrays{} within the accepted
\muAu{} \xray{} energy window.

The upstream and downstream target windows were thin
Form\-var foils (1\;cm diameter) of $10$\;$\mu$g/cm$^2$ at $\pHtwo=4$
and 16\;hPa, and $40$\;$\mu$g/cm$^2$ at 64\;hPa.
\Htwo{} gas with $<$\,3\;ppm impurities was continuously flushed
through the target (exchange rate 2/min).
Outgassing from walls and pipes resulted in
a total impurity concentration of order $10^{-4}$.
The measured \mup{} formation rate was $\sim$\,20\;s$^{-1}$.

A 20\;cm long cylindrical CsI(Tl) scintillator of 9.2\;cm inner and
19\;cm outer diameter was placed around the target. It had a high
probability to detect at least one of the \muAu{} \xrays{}.
The CsI(Tl) crystal was surrounded on all but the outer side
by an 8\;mm thick plastic scintillator acting as
anti-coincidence detector against charged particle background.
The fast light component of the plastic scintillator was distinguished
from the "slow" CsI light component (decay time $\sim\!1.5$\;\mus)
by pulse shape analysis.

%
%

%
For each incoming muon, the time difference
between the MCP signal and a CsI signal 
from a \muAu{} \xray{} was recorded.
The true \mup(\OneS) TOF, however, starts at the formation of a
\mup(\OneS).
Time calibration measurements were done with a few per cent of xenon
admixed to the \Htwo{} target gas, because at such a low Xe
concentration \muXe{} and \mup{} formation times are nearly
identical.
The \muXe{} \xrays{} were detected in the CsI(Tl) crystal, too.
Several Xe concentrations were used and the \mup(\OneS){} formation time
distribution was obtained by extrapolating to zero Xe concentration.
A careful analysis of the \muXe{} time spectra separated the time
distribution of direct \muXe-formation from the (delayed) transfer
reaction \mup\,+Xe\,$\rightarrow$\,$\muXe ^*$+p.
Small residual deviations not taken into account 
by the Xe calibration
(like the \mup{} cascade time and
discriminator threshold effects) were also considered.

%
\begin{figure}
  \includegraphics[width=\columnwidth]{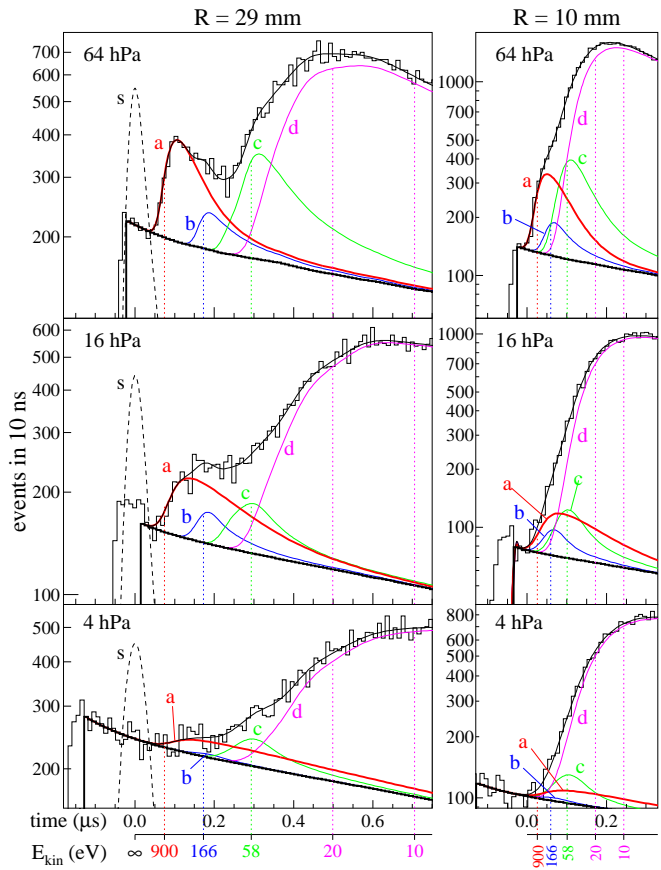}
    \caption{\label{fig:spectra} (color online). Early part of the
       \muAu{} time spectra measured at 64, 16, and 4\;hPa 
       (top to bottom)
       with the $R$\,=\,29\;mm (left)
       and $R$\,=\,10\;mm (right) targets.
       The corresponding \mup{} kinetic energies are also indicated.
       Component (a) at 0.9\;keV originates from non-radiative 
       \mup(\TwoS)-quenching
       and was convoluted with the fitted \TwoS-lifetime. 
       The thin solid line is the total fit function, composed of
       the measured background (thick black line), 
       the \TwoS{} component (a), the high energy
       Coulomb deexcitation lines with main components 
       (b) $n$\,=\,3\,$\rightarrow$\,2 (166\;eV) and
       (c) $n$\,=\,4\,$\rightarrow$\,3 (58\;eV), and
       a continuous distribution at energies below 60\;eV (d).
       The dashed curve (s) is the \mup{} formation time distribution
       obtained in the Xe calibration measurements.
      }
\end{figure}
%
Figure~\ref{fig:spectra} shows the \muAu{} time spectra measured at 64,
16, and 4\;hPa for both target radii, and
the \mup{} formation time distributions deduced from the
\muXe{} data.
The background was studied with the polypropylene target. It is
dominated by bremsstrahlung from muon-decay electrons, with an
additional component (depending on pressure) at times preceding \mup{}
formation, correlated to muons stopping in high-$Z$ materials near the
target.
All \muAu{} time spectra exhibit a wide distribution whose maximum
corresponds to \mup{} kinetic energies $E_{\rm kin}$ of $5 ...
20$\;eV.  Such energies result from Coulomb
deexcitations~\cite{BRAFIO78}
\mup$_n$+\Htwo\,$\rightarrow$\,\mup$_{n' < n}$\,+\,$E_{\rm kin}$\,+\,p\,+\,...
during the cascade, a well known effect for \mup{} and \pip{}
atoms~\cite{JENSEN02}.

The data measured in the $R$\,=\,29\;mm target, which is most
sensitive to high kinetic energies, show an unexpected peculiarity:
The 64\;hPa spectrum features an additional pronounced peak at
t\,$\approx\!100$\;ns.  This peak is also visible at 16\;hPa, but
almost disappears at 4\;hPa.  The corresponding kinetic energy is
about 0.9\;keV,
indicated in Fig.~\ref{fig:spectra} as dotted line at
0.074\;\mus{} which is the calculated TOF for 29\;mm, the shortest distance
from the beam center to the Au wall.

Such a high-energy \mup(\OneS)-component has never been reported before.
It can only originate from radiationless deexcitation of
\mup($n$\,=\,2) atoms in collisions with \Htwo{} molecules.
The $n$\,=\,2\,$\rightarrow$\,1 transition energy of 1.9\;keV 
is shared according to momentum conservation between
the formed \mup(\OneS) and a proton of the \Htwo{} molecule:
$$
 \mu {\prot}(n=2) + \Htwo{} \rightarrow  
   \mu {\prot}_{\OneS} {\rm (0.9\;keV)} + \prot{\rm (1\;keV)} + \dots \;\; .
$$
A detailed analysis of the measured time spectra was performed using a
set of Monte Carlo (MC) generated TOF spectra extending over
the whole expected energy range ($\sim\!10^{-2}$\,-\,$10^3$\;eV).
In the MC, the target and detector geometries,
axial and radial muon stop distributions, \xray{} detection
efficiencies, and scattering of \mup{} atoms on \Htwo{} molecules and
on the gold surface of the target cylinder
were taken into account~\cite{POHL01}.
Cross sections for \mup(\OneS)+\Htwo{} scattering calculated
by Adamczak~\cite{ADAM96a}
for laboratory kinetic energies below 9\;eV ($\approx$\,\Htwo{}
breakup energy) have been used.
Between 9\;eV and 1000\;eV  we applied those calculated by
Cohen~\cite{COH91b} for collisions with H-atoms. 
Coulomb deexcitation lines from low $n$-levels were included in the
fit function; the most important transitions are indicated in
Fig.~\ref{fig:spectra}.
A free fit gave only small amplitudes for Coulomb
deexcitation components with $\Delta n > 1$, in accordance with
calculations~\cite{BRAFIO78}.
Finally, each MC TOF spectrum was convoluted with
the measured \mup{} formation time distribution.
The fit function is a superposition of these spectra, with
their individual weights treated as free parameters. 
For each pressure, the two time spectra measured for $R$\,=\,10\;mm and
29\;mm were fitted simultaneously, providing a stringent consistency
test because the two time spectra are very different for the same
kinetic energy distribution.

An adequate fit of the data could not be obtained unless it was
assumed that the 0.9\;keV component was emitted with a non-zero
lifetime. Such an effect can only be caused by long-lived
\mup(\TwoS) atoms.
(Other collisional processes, like muon transfer to medium-$Z$ atoms
from gas impurities, would produce \xrays{} with energies below those
accepted for \muAu{}.)
For the data analysis a two-step process was adopted
where the \mup(\TwoS) atoms are first slowed down to low energies 
without significant quenching, and then collisionally
quenched at a constant rate \LamTSquench{}.
The 0.9\;keV MC time spectrum was therefore convoluted with an
exponential, with lifetime $\TauTSlong = (\LamTSquench + \lambda_\mu)^{-1}$  
and amplitude $\AmpTS{} = \EpsTSlong{} \, \LamTSquench{}$,
delayed by a time offset \OffTS{} for slowing down.
($\lambda_\mu$\,=\,$4.55\times 10^5$\;s$^{-1}$ is the muon decay rate.)

First the data sets at 64 and 16\;hPa were analyzed independently of
each other. (At 4\;hPa the 0.9\;keV component is too small to be
treated in an independent fit.)  Free fit parameters were the
\TwoS-quench time $\TauTSquench{} \equiv 1/\LamTSquench$, the
amplitude \AmpTS{}, the time offset \OffTS{}, and the amplitudes of
the Coulomb deexcitation components.
The results are given in Tab.~\ref{Table1}.\;
For \OffTS{} a value of  8\,$^{+18}_{-19}$\;ns was obtained at 16\;hPa,
and roughly 4 times less at 64\;hPa, i.e.\ much smaller than \TauTSquench{}.
%
\renewcommand{\arraystretch}{1.8}  
\begin{table}[b]
  \caption{\label{Table1} {\TwoS-quench time 
      $\TauTSquench$\,=\,$1/\LamTSquench (\approx \TauTSlong)$
      and population \EpsTSlong{} (per \mup{} atom) 
      of the long-lived \mup(\TwoS)-component, 
      obtained for the ``independent'' and ``combined'' fits of the time 
      spectra.
      For the combined fit we give the statistical and the systematic 
      errors, respectively.
      Only the statistical errors are given for the independent fits.
   } }
 \begin{tabular*}{8.6cm}{ @{} c @{\extracolsep\fill} c c c c @{} }\hline\hline
  \pHtwo & \multicolumn{2}{c}{independent fit}   
                    & \multicolumn{2}{c}{combined fit} \\
   (hPa) & \TauTSquench(ns) & \EpsTSlong{}\,(\%) 
         & \TauTSquench(ns) & \EpsTSlong{}\,(\%) \\ \hline
     64  & $        30\,^{+23} _{-8}$  &   
           $       1.6\,^{+0.9} _{-0.3}$    &
           $      30.4\,^{+19.7\;+8.4} _{-7.3\;\;\,-6.3}$  & 
           $      1.63\,^{+0.73\;+0.35} _{-0.20\;-0.15}$  \\
     16  & $     124  \,^{+542} _{-54}$  &   
           $       1.1\,^{+2.9} _{-0.4}      $   &
           $         4 \times  $ &
           $      1.07\,^{+0.41\;+0.16} _{-0.16\;-0.14}$  \\
  \,\;4  &              -      &     -      &   
           $         16 \times  $ &
           $      1.25\,^{+0.69\;+0.29} _{-0.25\;-0.23}$  \\ \hline\hline
 \end{tabular*}
\end{table}
%

Next, \TauTSquench{} was set to be proportional to 1/\pHtwo{}, because
non-linear effects (like three-body collisions) can be excluded at our
conditions. A combined fit of the data at all three pressures gave nearly
the same values as the independent fit (Tab.~\ref{Table1}).
The corresponding quench rate at room temperature is
$$
  \LamTSquench  =  5.1^{+2.4}_{-2.1} \times 10^5\;{\rm s}^{-1}
  ~ \times ~ \pHtwo{}\;{\rm [hPa]} 
$$
or, normalized to liquid hydrogen atom density 
(LHD\,=\,$4.25 \times 10^{22}$\;atoms/cm$^3$), 
\begin{equation} 
   \LamTSquench({\rm LHD}) = 4.4\,^{+2.1}_{-1.8} 
                              \times 10^{11}\;{\rm s}^{-1} .
 \label{eq:result}
\end{equation}
The errors result from the quadratic sum of statistical and systematic
errors.
The combined fit gave a reduced \ChiSq{} of 1.030 (\ChiSq\,=\,529.5 for
514 degrees of freedom).  \ChiSq\,=\,559.2 resulted for \TauTSquench(64\,hPa)
fixed to 9\;ns, i.e.\ such a short \TwoS-lifetime is
statistically excluded at the 5$\sigma$ confidence level.

Various sources of systematic errors were investigated:
\mup{} formation time, scattering on H$_2$ and gold,
radial size of the muon stop distribution, and background shape.
Uncertainties in the formation time gave a contribution of
$^{+8.0}_{-4.7}$\;ns to \TauTSquench(64\;hPa), 
H$_2$-scattering one of $^{+1.0}_{-4.0}$\;ns,
whereas all other errors were considerably smaller. 
Gas impurities modified \TauTSquench(64\;hPa) by $<$\,0.1\;ns.
The resulting total systematic errors are given in Tab.~\ref{Table1}.

The large value found for \LamTSquench{} can be attributed to the
formation of an {\it excited} muonic molecular ion in a resonantly
enhanced reaction with \Htwo{}~\cite{FROEL93,WALL96}
$$
 \mu {\prot}(\TwoS) + \Htwo{} \rightarrow \ppmupee \;,
$$
similar to the well-known formation of ground state muonic molecules in
muon catalyzed fusion~\cite{SCRI93}.
There is a series of resonant states just below the
\mup($n$\,=\,2)\,+\,p dissociation limit~\cite{LIND03}; 
rovibrational excitations of the electronic molecule absorb the small
excess energy from the formation of the \ppmu{} ``pseudonucleus''.
A subsequent Auger decay to states with $\ge 16$\;eV binding energy
stabilizes the \ppmustar{} against back decay. 
Mixing of the excited \ppmustar{} with the repulsive \ppmu{} ground
state finally leads to \mup(\OneS) atoms with 0.9\;keV kinetic energy.

A theoretical estimate of the molecular formation rate~\cite{WALL01}
gave $\sim\!5 \times 10^{10}$\;s$^{-1}$ at LHD,
ten times smaller than our result (Eq.~(\ref{eq:result})).
The calculation neglected the correlated motion of the three
interacting atoms which is important in low-energy \mup(\TwoS)
collisions because of its long-range polarization potential. The
triatomic molecule has a much denser vibrational-state spectrum with
more ways to exchange excitation energy, leading to longer effective
lifetimes of the compound system~\cite{TAQQU88} and correspondingly
higher molecular formation rates.

The populations \EpsTSlong{} resulting at 16 and 4\;hPa agree with the
values of $(1.16\pm 0.12)$\,\% and $(1.03\pm 0.09)$\,\%, respectively,
evaluated for the \mup(\TwoS) that slow down to below
0.3\;eV~\cite{POHL01}.
This means that there is little room for radiative
\TwoS\,$\rightarrow$\,\OneS{} deexcitation processes
and explains why no delayed 1.9\;keV \xrays{} were found
in previous experiments~\cite{ANDE77,EGAN81,BOEC82}.
The dominance of non-radiative \TwoS-decay resulting from our
measurements is in accord with recent calculations
of the radiative and non-radiative decay rates of \ppmustar{} 
molecules~\cite{LIND03,KILIC04}.

In summary, we have measured that $\sim\!1$\,\% of all muons stopped
in \Htwo{} gas between 4 and 64\;hPa form long-lived \mup(\TwoS)
atoms that decay non-radiatively
into \mup(\OneS) atoms of 0.9\;keV kinetic energy.
For the current \mup{} Lamb shift experiment, a pressure of $\sim\!1$\;hPa is
needed where both \TwoS-population ($\sim\!1\,\%$) and lifetime
($\TauTSlong = 1.04\,^{+0.29}_{-0.21}$\;\mus{} at 1\;hPa) meet the
experimental requirements~\cite{POHL05}.
First measurements at a new low-energy muon beam at PSI
showed that enough \mup{} atoms can be formed in a small gas
volume at such pressures.

We thank G.\;Llos\'a and Ch.\;Maierl for their help during data taking,
A.\;Adamczak, J.S.\;Cohen, and V.S.\;Melezhik for providing
us with \mup(\OneS) scattering cross sections,  
T.\;Jensen, S.\;Romanov, and J.\;Wallenius for fruitful discussions, 
M.\;Aigner and H.\;Angerer for technical support, 
P.\;Maier-Komor and K.\;Nacke for manufacturing of ultra-thin C-foils, 
and the PSI staff for providing us with an excellent beam.

\end{document}